\documentclass[12pt]{article}
\usepackage{epsf}
\begin{document}

\title{Two-Dimensional Coulomb Gas \\ on an Elliptic Annulus}

\author{Taro Nagao} 

\date{} 

\maketitle 

\begin{center} 
\it Graduate School of Mathematics, Nagoya University, Chikusa-ku,
\\  Nagoya 464-8602, Japan \\
\end{center}

\begin{center}
E-mail: nagao@math.nagoya-u.ac.jp
\end{center}

\begin{abstract}
It is well-known that two-dimensional Coulomb gases at 
a special inverse temperature $\beta = 2$ can be analyzed 
by using the orthogonal polynomial method borrowed from the 
theory of random matrices. In this paper, such Coulomb gas molecules  
are  studied when they are distributed on an elliptic annulus, and the 
asymptotic forms of the molecule correlation functions in the thermodynamic 
limit are evaluated. For that purpose, two-dimensional orthogonality 
relations of the Chebyshev polynomials on an elliptic annulus are utilized. 

\end{abstract}


\medskip


\newpage

\section{Introduction}
\setcounter{equation}{0}
\renewcommand{\theequation}{1.\arabic{equation}}

Let us consider two-dimensional one-component 
Coulomb gas systems with Hamiltonians
\begin{equation}
{\cal H} = \sum_{j=1}^N {\cal V}(z_j) - \sum_{j<l}^N \log|z_j - z_l|.
\end{equation}
Here the locations $(x_j,y_j)$ of logarithmically interacting 
$N$ molecules are denoted by complex numbers $z_j = x_j + i y_j$ ($j = 
1,2,\cdots,N$) on the two dimensional complex plane ${\bf C}$. 
The one-particle potential functions are given by ${\cal V}(z)$ equated with ${\cal V}({\bar z})$ , 
where $z = x + i y$ and ${\bar z} = x - i y$ with real numbers $x$ and $y$.  Such logarithmically 
interacting molecule systems on a two-dimensional (2D) surface are in general called 2D 
Coulomb gases, because 2D Coulomb interaction of charged molecules is logarithmic\cite{chafai}. In this paper, one-component gases with only repulsive interactions are studied. 
\par
Coulomb gas systems are  important and useful in physics, not only from the viewpoint of 
classical statistical mechanics of charged particles, but also from the applications to 
random matrix models of disordered or chaotic quantum systems. Many famous random 
matrix models can be derived as various limits of 2D Coulomb gases. Therefore we can 
hopefully develop unified ways to study those models and observe the transitions between 
them. This gives a motivation to make research on Coulomb gas systems. 
\par
The equilibrium probability density function of the 2D Coulomb gas systems
at an inverse temperature $\beta = 2$ is known to be
\begin{equation}
\label{pdf}
P(z_1,z_2,\cdots,z_N) \propto e^{- \beta {\cal H}} 
= \prod_{j=1}^N w_{\bf C}(z_j) \prod_{j<l}^N |z_j - z_l|^2
\end{equation}
with a one-molecule weight function $w_{\bf C}(z) = e^{- 2 {\cal V}(z)}$ and 
an integration measure $\prod_{j=1}^N dz_j = 
\prod_{j=1}^N dx_j dy_j$. Here $P(z_1,z_2,\cdots,z_N)$ is normalized as
\begin{equation}
\int_{\bf C}  dz_1 \int_{\bf C}
dz_2 \cdots \int_{\bf C} dz_N \ P(z_1,z_2,\cdots,z_N) = 1, 
\end{equation}
and we define the $k$-molecule correlation functions
\begin{equation}
\rho(z_1,z_2,\cdots,z_k) = \frac{N!}{(N-k)!} \int_{\bf C}  dz_{k+1} \int_{\bf C}
dz_{k + 2} \cdots \int_{\bf C} dz_N \ P(z_1,z_2,\cdots,z_N),
\end{equation}
which are typical observable physical quantities. 
\par
As is well-known, the probability density function 
(\ref{pdf}) can be analyzed by using the corresponding orthogonal 
polynomials on the complex plane.  This fact originated in 
Ginibre's seminal work\cite{ginibre} on random matrices and then 
extended to general Coulomb gas systems of the above 
form\cite{pjf1,pjf2}. Suppose that $M_n(z) = z^n +({\rm lower 
\ order \ terms})$ with non-negative integers $n$ are the monic orthogonal polynomials 
satisfying the orthogonality relation 
\begin{equation}
\int_{\bf C} \sqrt{w_{\bf C}({\bar z}) w_{\bf C}(z)} M_m({\bar z}) M_n(z) dz =  h_n \delta_{mn}
\end{equation}
with $m,n = 0,1,2,\cdots$.  Then the $k$-molecule correlation functions can be written in 
determinant forms\cite{KS}
\begin{equation}
\label{rhodet}
\rho(z_1,z_2,\cdots,z_k) = {\rm det}\left[ K_N(z_j, z_{\ell}) \right]_{j,\ell = 1,2,\cdots,k},
\end{equation}
where
\begin{equation}
K_N(z_j,z_{\ell}) = \sqrt{w_{\bf C}(z_j) w_{\bf C}({\bar z_{\ell}})} \sum_{n = 0}^{N-1} \frac{1}{h_n} M_n(z_j) M_n({\bar z_{\ell}}).
\end{equation}
Therefore, in order to see the asymptotic behavior of the $k$-molecule 
correlation functions in the thermodynamic limit $N \rightarrow \infty$, we only need to 
evaluate the kernel function $K_N(z_1,z_2)$.
\par
As explained above, in the study of a 2D Coulomb gas system at 
an inverse temperature $\beta = 2$, the corresponding orthogonal polynomials 
$M_n(z)$ can be conveniently utilized. Radially symmetric cases, in 
which $w_{\bf C}(z) $ only depends on the radial variable $|z|$, 
are particularly simple, because the corresponding orthogonal polynomials 
are monomials $z^n$. Important classes of non-hermitian random 
matrix problems,  such as truncations of unitary matrices\cite{ZS}, gap probabilities\cite{APS}, induced ensembles\cite{FBKSZ} and products of random matrices\cite{AB} can be treated by using monomials, because 
they are reduced to the analysis of radially symmetric Coulomb gas systems. Monomials have been powerful tools in the analysis of non-hermitian random 
matrices.
\par
Why are monomials so useful? One of the reasons is that the 
orthogonality of monomials comes from the integration only over the angular 
variable. Then we can freely choose a radial weight function 
depending only on the radial variable. For that reason, we can generate a 
large class of Coulomb gas systems, which can be treated  by means of monomials. 
For example, Coulomb gases on an annulus  with a hole at the central domain 
can be studied by using monomials.
\par
In radially asymmetric cases, on the other hand, we need to use more 
general orthogonal polynomials. The examples are the Coulomb gas 
systems related to Gaussian ensembles of weakly non-hermitian random matrices\cite{FGIL,osborn, akemann}. The corresponding orthogonal polynomials are the  Hermite and Laguerre 
polynomials. More recently,  a relation between Coulomb gas systems on 
the domain inside an ellipse and the Gegenbauer polynomials was found, 
and the orthogonality of the Gegenbauer polynomials was used to 
analyze such Coulomb gases\cite{NAKP,ANPV}. The Gaussian random matrix ensembles
related to the Hermitian polynomials are regarded as asymptotic 
limits of such gases.  However, the orthogonality of the Gegenbauer 
polynomials in general holds for  the integration over both angular 
and radial variables. Therefore,  the Coulomb gas systems related to the 
Gegenbauer polynomials are limited,  because we cannot freely choose 
the radial weight function.  In particular, extensions to annulus cases 
are not straightforward.
\par
In spite of the above situation, there still exist Coulomb gas systems, which are radially 
asymmetric and related to the Chebyshev polynomials, for which 
the orthogonality comes from the integration only over the angular 
variable. Then we can freely choose a radial weight 
function, and generate a large class of Coulomb gas systems. 
Such an exceptional property of the Chebyshev polynomials has been known in mathematical 
literature for a long time\cite{walsh1,szego1,walsh2,PH,MH}. In this paper 
we employ this useful property to study Coulomb gas systems on an elliptic annulus. 
\par
We are in a position to explain a couple of physical motivations to consider annulus 
systems. One motivation is to observe the transition between 2D and 
one-dimensional (1D) systems. There are well-known 1D log-gases, 
resulting from random matrix models such as circular and Gaussian ensembles.  
As discussed in \cite{chafai}, these 1D log-gases are regarded as 2D 
Coulomb gases on 1D lines or curves embedded into a 2D surface.  However, 
from a viewpoint of statistical mechanics of charged particles, a quasi 1D system 
is more realistic than such a strictly 1D system. A quasi 1D system has a  small but 
non-zero width on the 2D surface, and exhibits physical properties between 1D and fully 2D systems.  This motivates the study 
of thin annulus systems which are quasi 1D.  In this paper, universal asymptotic 
forms of the molecule correlation functions are derived for such quasi 1D systems, which include radially asymmetric cases, and limits to strictly 
1D systems are discussed.  
\par
Another physical motivation comes from the interest on the gap probabilities leading to  
the distribution  functions of ordered molecules\cite{APS}, such as the density of the 
molecule with the smallest radial variable. For example, let us suppose that $N$ 
molecules are distributed on a disc $|z| \leq v$. We fictitiously set the corresponding annulus 
weight function as
\begin{equation}
w_{\bf C}(z) = \left\{ \begin{array}{ll}
1, &  R \leq |z| \leq v,  \\
0, & {\rm otherwise} \end{array} \right.
\end{equation}
($0 < R <v$) and calculate the one-molecule correlation function $\rho(z_1)$.  Then 
the density of the molecule at a location $z$ with the smallest radial variable $|z|=R$ is 
given by $\displaystyle \lim_{|z| \downarrow R}\rho(z)$. In a similar way, by using annulus weight functions, we can also evaluate various distribution functions of 
ordered molecules. 
\par
Let us now consider an elliptic annulus
\begin{equation}
\label{elliptic-annulus}
{\cal A} = \left\{ z \mid (x/a_v)^2 + (y/b_v)^2  \leq 1 \ {\rm and} \ (x/a_R)^2  + (y/b_R)^2  \geq 1 \right\}
\end{equation}
on the complex plane. Here $z = x + i y$ ($x$ and $y$ are real), 
\begin{equation}
a_v = \frac{1}{2} \left( v + \frac{1}{v} \right), \ \ \ 
b_v = \frac{1}{2} \left( v - \frac{1}{v} \right),
\end{equation}
\begin{equation}
a_R = \frac{1}{2} \left( R + \frac{1}{R} \right), \ \ \ 
b_R = \frac{1}{2} \left( R - \frac{1}{R} \right)
\end{equation}
and $1 < R < v$. In this paper, we are interested in the Coulomb gas systems confined on 
the elliptic annulus ${\cal A}$. Namely the one-molecule weight function $w_{\bf C}(z)$ 
has a form
\begin{equation}
w_{\bf C}(z) = \left\{ \begin{array}{ll}
w_{\cal A}(z), & z \in {\cal A}, \\
0, & z \notin {\cal A}. \end{array} \right.
\end{equation}
Now we moreover suppose that the weight function $w_{\cal A}(z)$ on the annulus has a flat form
\begin{equation}
\label{waz1}
w_{\cal A}(z) = 1
\end{equation}
or one of the forms
\begin{equation}
\label{waz2}
w_{\cal A}(z) = \frac{1}{| 1 - z^2 |}, \ \ \frac{1}{|1 - z|}  \ {\rm and} \ \frac{1}{|1 + z|}.
\end{equation} 
In these cases, we find that the corresponding orthogonal polynomials on the 
elliptic annulus are proportional to the Chebyshev polynomials. Then we can 
evaluate the asymptotic formulas of the kernel functions in the limit $N \rightarrow \infty$. 
\par
This paper is organized as follows. In \S 2, the orthogonality of the Chebyshev 
polynomials on the elliptic annulus is proved and used to obtain kernel function 
formulas for  a finite $N$. In \S 3, $N \rightarrow \infty$ limits of the kernel functions 
are evaluated, when the elliptic annulus is contained in the neighborhood of the 
outer edge, and when it is in the neighborhood of the real interval $[-1,1]$.  
As discussed above, one of the physical motivations focuses on these quasi 1D cases. 
In Appendices, conventional Coulomb gas results on a disc and on a real 1D 
interval are derived for comparison. 

\section{Chebyshev polynomials}
\setcounter{equation}{0}
\renewcommand{\theequation}{2.\arabic{equation}}

For a complex coordinate $z = x + i y$ ($x$ and $y$ are real)  which is 
not on the real interval $[-1,1]$,  we introduce a complex 
parameter $\omega$ as
\begin{equation}
z = \frac{1}{2} \left( \omega + \frac{1}{\omega} \right)
\end{equation}
with $|\omega| > 1$. Then the Chebyshev polynomials 
are defined in the following way.  A Chebyshev polynomial 
with a non-negative integer index $n$ is an $n$-th degree 
polynomial of $z$.

\par\medskip\noindent
(1) Chebyshev polynomials of the first kind
\begin{equation}
T_n(z) = \frac{1}{2} \left( \omega^n + \omega^{-n} \right).
\end{equation}
The corresponding monic orthogonal polynomials $M_n(z)$ are 
\begin{equation}
M_n^{(1)}(z) = \left\{ \begin{array}{ll}
1, & n = 0, \\  \displaystyle
\frac{1}{2^{n-1}} T_n(z), & \displaystyle n > 0. \end{array} \right.
\end{equation}
\par\medskip\noindent
(2) Chebyshev polynomials of the second kind
\begin{equation}
U_n(z) = \frac{\omega^{n + 1} - \omega^{-n-1}}{\omega - \omega^{-1}}
\end{equation}
with the monic orthogonal polynomials $M_n^{(2)}(z) = (1/2^n) U_n(z)$. 
\par\medskip\noindent
(3) Chebyshev polynomials of the third kind
\begin{equation}
V_n(z) = \frac{\omega^{n + 1} + \omega^{-n}}{\omega + 1}
\end{equation}
with the monic orthogonal polynomials $M_n^{(3)}(z) = (1/2^n) V_n(z)$. 
\par\medskip\noindent
(4) Chebyshev polynomials of the fourth kind
\begin{equation}
W_n(z) = \frac{\omega^{n + 1} - \omega^{-n}}{\omega - 1}
\end{equation}
with the monic orthogonal polynomials $M_n^{(4)}(z) = (1/2^n) W_n(z)$. 
\par
The Chebyshev polynomials are known to satisfy orthogonality relations\cite{walsh1, szego1, walsh2, PH, MH} 
on the elliptic region
\begin{equation}
{\cal E} = \left\{ z = x + i y \mid (x/a_v)^2 + (y/b_v)^2  \leq 1 \right\}
\end{equation}
with the weight functions (\ref{waz1}) and (\ref{waz2}). We can readily extend them to the 
orthogonality relations on the elliptic annulus ${\cal A}$. 

\subsection{Model I}
Let us first consider the Chebyshev polynomials of the first kind 
with the Model I weight function
\begin{equation}
\label{wa1}
w_{\cal A}(z) = \frac{1}{| 1 - z^2 |}.
\end{equation} 
The complex parameter $\omega$ can be written in terms of the 
polar coordinates $(r,\theta)$ as
\begin{equation}
\omega = r e^{i \theta}, \ \ \ r > 1, \ \ \ \theta \ {\rm real}.
\end{equation}
In an elliptic system, $r$ plays the role of a radial variable, and $\theta$ plays the role 
of an angular variable. It follows from $z = x + i y$ that
\begin{eqnarray}
& & \frac{\partial x}{\partial r} =  \frac{1}{2 r} \left( r - \frac{1}{r} \right) \cos\theta,  \ \  
\frac{\partial x}{\partial \theta} =  - \frac{1}{2} \left( r + \frac{1}{r} \right) \sin\theta,  \nonumber \\ 
& & \frac{\partial y}{\partial r} =  \frac{1}{2 r} \left( r + \frac{1}{r} \right) \sin\theta, \ \ 
\frac{\partial y}{\partial \theta} = \frac{1}{2} \left( r - \frac{1}{r} \right) \cos\theta
\end{eqnarray}
and
\begin{equation}
\label{jacobian}
\frac{\partial(x,y)}{\partial(r,\theta)} = \frac{\partial x}{\partial r} \frac{\partial y}{\partial \theta} - 
\frac{\partial x}{\partial \theta} \frac{\partial y}{\partial r} = \frac{1}{4 r} \left| \omega - \frac{1}{\omega} \right|^2 
= \frac{|1 - z^2|}{r}.
\end{equation}
When $r = v$, 
\begin{eqnarray}
z & = & \frac{1}{2} \left( \omega + \frac{1}{\omega} \right) 
= \frac{1}{2} \left( v e^{i \theta} + \frac{1}{v} e^{- i \theta} \right) \nonumber \\ 
& = & a_v \cos\theta  + i b_v \sin\theta
\end{eqnarray}
is on the boundary of the elliptic region ${\cal E}$.  
\par
Let us now take an arbitrary function $f(r)$ of $r$. If relevant integrals exist, the 2D integral 
\begin{equation}
I_{mn} = \int_{\cal E} dx dy \frac{f(r)}{| 1 - z^2 |} 
M^{(1)}_m({\bar z})  
M^{(1)}_n(z) 
\end{equation}
over the elliptic region ${\cal E}$ can be evaluated as
\begin{eqnarray}
I_{mn} & = & 2^{2 - m - n}  \int_1^v dr \int_0^{2 \pi} d\theta  \frac{f(r)}{r} 
T_m({\bar z}) T_n(z) \nonumber \\ 
& = & 2^{- m - n}  \int_1^v dr   \int_0^{2 \pi} d\theta \frac{f(r)}{r} 
({\bar \omega}^m + {\bar \omega}^{-m}) (\omega^n + \omega^{-n})  \nonumber \\ 
& = & 2^{- m - n}  \int_1^v dr    \frac{f(r)}{r}  \int_0^{2 \pi} d\theta  \nonumber \\ & & \times
\left(r^{m+n} e^{i (n - m) \theta} + r^{m-n} e^{- i (m + n) \theta} + r^{n - m} e^{i (m+n) \theta} 
+ r^{-m-n} e^{i (m - n)\theta} \right)
 \nonumber \\ 
& = & h_n \delta_{mn}
\end{eqnarray}
for $m,n > 0$. In general, $I_{mn} = h_n \delta_{mn}$ for $m,n \geq 0$ with
\begin{equation}
h_n = \frac{\pi}{2^{2 n - 1}} \int_1^v dr f(r) \left(r^{2 n-1} + r^{-2 n - 1} \right), \ \ \ n > 0
\end{equation}
and
\begin{equation}
h_0 = 2 \pi \int_1^r dr \frac{f(r)}{r}.
\end{equation}
Therefore $M^{(1)}_n(z) $ are monic orthogonal polynomials with 
a weight function
\begin{equation}
w_{\bf C}(z) = \left\{ \begin{array}{ll}
\displaystyle \frac{f(r) }{| 1 - z^2 |}, & z \in {\cal E}, \\
0, & z \notin {\cal E}  \end{array} \right.
\end{equation}
on the complex plane. 
\par
Now we take a special form of $f(r)$ as
\begin{equation}
\label{fr}
f(r) = \left\{ \begin{array}{ll} 1,  & r \geq R, \\
0,  & r < R \end{array} \right.
\end{equation}
with $1 < R < v$.  This special form means that an inner hard wall is introduced at $r = R$. As explained in Introduction, when one considers the gap probabilities\cite{APS},  a fictitious inner 
hard wall can be useful. When $r = R$, 
\begin{eqnarray}
z & = & \frac{1}{2} \left( \omega + \frac{1}{\omega} \right) 
= \frac{1}{2} \left( R e^{i \theta} + \frac{1}{R} e^{- i \theta} \right) \nonumber \\ 
& = & a_R \cos\theta  + i b_R \sin\theta
\end{eqnarray}
is on the inner boundary of the elliptic annulus  ${\cal A}$ defined in (\ref{elliptic-annulus}). Therefore the Coulomb gas system is confined on ${\cal A}$. Then the orthogonality relation
\begin{equation}
\int_{\cal A} dx dy \frac{1}{| 1 - z^2 |}  M^{(1)}_m({\bar z})  M^{(1)}_n(z) = h^{(1)}_n \delta_{mn}, \ \ \ m,n \geq 0
\end{equation}
with
\begin{eqnarray}
h^{(1)}_n & = & \frac{\pi}{2^{2 n - 1}} \int_R^v dr  (r^{2 n-1} + r^{-2 n - 1})  \nonumber \\ 
& = & \frac{\pi}{2^{2 n} n} \left( v^{2 n} - v^{-2 n} - R^{2 n} + R^{-2 n} \right), \ \ \ n > 0
\end{eqnarray}
and
\begin{equation}
h^{(1)}_0 = 2 \pi \int_R^v dr \frac{1}{r} = 2 \pi \left( \log v - \log R \right)
\end{equation}
holds. It follows that the kernel function $K_N(z_1,z_2)$ with $N > 1$ is equal to
\begin{eqnarray}
\label{k1}
& & K_N^{(1)}(z_1,z_2) = \frac{1}{\displaystyle \sqrt{|1 - (z_1)^2| |1 - (z_2)^2|}} 
\sum_{n=0}^{N-1} \frac{1}{h_n^{(1)}} M_n^{(1)} (z_1) M_n^{(1)}({\bar z_2}) \nonumber \\ 
& = &  \frac{1}{\displaystyle \sqrt{|1 - (z_1)^2| |1 - (z_2)^2|}}  \nonumber \\ 
& & \times\left\{  \frac{1}{2 \pi \log(v/R)} + \frac{1}{\pi}  \sum_{n=1}^{N-1} n 
\frac{((\omega_1)^n + (\omega_1)^{-n}) ( ({\bar \omega_2})^n + ({\bar \omega_2})^{-n})}{
v^{2 n} - v^{- 2n} - R^{2 n} + R^{- 2 n}}
\right\}, \nonumber \\ 
\end{eqnarray}
where
\begin{equation}
z_j = \frac{1}{2} \left( \omega_j + \frac{1}{\omega_j} \right)
\end{equation}
with $|\omega_j| > 1$ and $j = 1,2$.

\subsection{ Model II}
Next we consider the Chebyshev polynomials of the second kind corresponding to the flat Model II 
weight function
\begin{equation}
\label{wa2}
w_{\cal A}(z) = 1.
\end{equation}
As before, the orthogonality relation for the Chebyshev polynomials of the second kind on the 
elliptic region ${\cal E}$ can be derived as
\begin{equation}
\int_{\cal E} dx dy f(r) M_m^{(2)}({\bar z}) M_n^{(2)}(z) = h_n \delta_{mn}, \ \ \ m,n \geq 0,
\end{equation}
where 
\begin{equation}
h_n = \frac{\pi}{2^{2 n + 1}} \int_1^v dr f(r) \left(r^{2 n + 1} + r^{-2 n - 3} \right), \ \ n \geq 0.
\end{equation}
Focusing on  the special case (\ref{fr}), we arrive at the orthogonality relation on the annulus ${\cal A}$
\begin{equation}
\int_{\cal A} dx dy  M_m^{(2)}({\bar z}) M_n^{(2)}(z) = h^{(2)}_n \delta_{mn}, \ \ m,n \geq 0,
\end{equation}
where
\begin{equation}
h^{(2)}_n = \frac{\pi}{2^{2 n + 2} (n + 1)}  \left( v^{2 n + 2} - v^{-2 n - 2} - R^{2 n + 2} + R^{-2 n - 2} 
\right), \ \ n \geq 0.
\end{equation}
Therefore the kernel function $K_N(z_1,z_2)$ is equal to
\begin{eqnarray}
\label{k2}
& & K_N^{(2)}(z_1,z_2) = 
\sum_{n=0}^{N-1} \frac{1}{h_n^{(2)}} M_n^{(2)} (z_1) M_n^{(2)}({\bar z_2}) \nonumber \\ 
& = &  \frac{4}{\pi}  \frac{1}{(\omega_1 - \omega_1^{-1}) ({\bar \omega_2} - {\bar \omega_2}^{-1})}
\sum_{n=1}^N n \frac{((\omega_1)^n - (\omega_1)^{-n}) ( ({\bar \omega_2})^n - ({\bar \omega_2})^{-n})}{
v^{2 n} - v^{- 2n} - R^{2 n} + R^{- 2 n}}. \nonumber \\ 
\end{eqnarray}

\subsection{Model III} 
The Chebyshev polynomials of the third kind correspond to the Model III weight function
\begin{equation}
\label{wa3}
w_{\cal A}(z) = \frac{1}{|1 - z|}.
\end{equation}
Following the previous procedures, we can similarly obtain the orthogonality relation for the 
Chebyshev polynomials of the third kind as
\begin{equation}
\int_{\cal E} dx dy \frac{1}{|1 - z|} f(r) M_m^{(3)}({\bar z}) M_n^{(3)}(z) = h_n \delta_{mn}, \ \ \ m,n \geq 0,
\end{equation}
where 
\begin{equation}
h_n = \frac{\pi}{2^{2 n}} \int_1^v dr f(r) \left(r^{2 n} + r^{-2 n - 2} \right), \ \ n \geq 0.
\end{equation}
We restrict ourselves to the special case (\ref{fr}) and find the orthogonality relation
\begin{equation}
\int_{\cal A} dx dy  \frac{1}{|1 - z|} M_m^{(3)}({\bar z}) M_n^{(3)}(z) = h^{(3)}_n \delta_{mn}, \ \ m,n \geq 0,
\end{equation}
where
\begin{equation}
h^{(3)}_n = \frac{\pi}{2^{2 n} (2 n + 1)}  \left( v^{2 n + 1} - v^{-2 n - 1} - R^{2 n + 1} + R^{-2 n - 1} 
\right), \ \ n \geq 0.
\end{equation}
Then the kernel function $K_N(z_1,z_2)$ is equated with
\begin{eqnarray}
\label{k3}
& & K_N^{(3)}(z_1,z_2) = \frac{1}{\sqrt{|1 - z_1| | 1 - z_2|}} 
\sum_{n=0}^{N-1} \frac{1}{h_n^{(3)}} M_n^{(3)} (z_1) M_n^{(3)}({\bar z_2}) \nonumber \\ 
& = &  \frac{1}{\pi}  \frac{1}{(\omega_1 + 1) ({\bar \omega_2} + 1)} 
\frac{1}{\sqrt{|1 - z_1| | 1 - z_2|}} \nonumber \\ 
& & \times \sum_{n=0}^{N-1} (2  n + 1)  \frac{((\omega_1)^{n+1} + (\omega_1)^{-n}) ( ({\bar \omega_2})^{n+1} 
+ ({\bar \omega_2})^{-n})}{
v^{2 n+1} - v^{- 2n-1} - R^{2 n+1} + R^{- 2 n-1}}. 
\end{eqnarray}

\subsection{Model IV}
Finally,  for the Chebyshev polynomials of the fourth kind associated with the 
Model IV weight function
\begin{equation}
\label{wa4}
w_{\cal A}(z) = \frac{1}{|1 + z|}, 
\end{equation}
the orthogonality relation turns out to be
\begin{equation}
\int_{\cal E} dx dy \frac{1}{|1 + z|} f(r) M_m^{(4)}({\bar z}) M_n^{(4)}(z) = h_n \delta_{mn}, \ \ \ m,n \geq 0,
\end{equation}
where 
\begin{equation}
h_n = \frac{\pi}{2^{2 n}} \int_1^v dr f(r) \left(r^{2 n} + r^{-2 n - 2} \right), \ \ n \geq 0.
\end{equation}
In the special case (\ref{fr}) , it is reduced to be
\begin{equation}
\int_{\cal A} dx dy  \frac{1}{|1 + z|} M_m^{(4)}({\bar z}) M_n^{(4)}(z) = h^{(4)}_n \delta_{mn}, \ \ m,n \geq 0,
\end{equation}
where
\begin{equation}
h^{(4)}_n = \frac{\pi}{2^{2 n} (2 n + 1)}  \left( v^{2 n + 1} - v^{-2 n - 1} - R^{2 n + 1} + R^{-2 n - 1} 
\right), \ \ n \geq 0.
\end{equation}
We then find the equality of the kernel functions $K_N(z_1,z_2)$ and
\begin{eqnarray}
\label{k4}
& & K_N^{(4)}(z_1,z_2) = \frac{1}{\sqrt{|1 + z_1| | 1 + z_2|}} 
\sum_{n=0}^{N-1} \frac{1}{h_n^{(4)}} M_n^{(4)} (z_1) M_n^{(4)}({\bar z_2}) \nonumber \\ 
& = &  \frac{1}{\pi}  \frac{1}{(\omega_1 - 1) ({\bar \omega_2} - 1)} 
\frac{1}{\sqrt{|1 + z_1| | 1 + z_2|}} \nonumber \\ 
& & \times \sum_{n=0}^{N-1} (2  n + 1)  \frac{((\omega_1)^{n+1} - (\omega_1)^{-n}) ( ({\bar \omega_2})^{n+1} 
- ({\bar \omega_2})^{-n})}{
v^{2 n+1} - v^{- 2n-1} - R^{2 n+1} + R^{- 2 n-1}}. 
\end{eqnarray}

\section{Asymptotic forms}
\setcounter{equation}{0}
\renewcommand{\theequation}{3.\arabic{equation}}

In this section, we discuss the asymptotic forms of the kernel functions in 
the thermodynamic limit $N \rightarrow \infty$. 

\subsection{Model I}
Our first case is equipped with the Model I weight function (\ref{wa1}), for 
which the kernel function $K_N^{(1)}(z_1,z_2)$ is given  
in (\ref{k1}). Now we introduce the polar coordinates $r_1,r_2 > 1$ and 
real $\theta_1,\theta_2$ as
\begin{equation}
\label{polar}
\omega_j = r_j e^{i \theta_j}, \ \ \ j = 1,2.
\end{equation}
\par
In a quasi 1D case, when the elliptic annulus is contained in  the neighborhood of the 
outer edge $r_1,r_2 = v > 1$, we introduce scaling parameters 
$t_1$, $t_2$, $\phi_1$, $\phi_2$ and $T$ as
\begin{equation}
\label{scaling1}
r_1 = v  \left( 1 - \frac{t_1}{N} \right), \ \ \ r_2 = v \left( 1 - \frac{t_2}{N} \right),
\end{equation}
\begin{equation}
\label{scaling2}
\theta_1 = \psi + \frac{\phi_1}{N}, \ \ \  \theta_2 = \psi + \frac{\phi_2}{N}
\end{equation}
and
\begin{equation}
\label{scaling3}
R = v \left( 1 - \frac{T}{N} \right),
\end{equation}
where $0 < t_1,t_2< T$ and $0 \leq  \psi < 2 \pi$.  Thus the angular 
coordinates $\theta_1,\theta_2$ are in the neighborhood of a 
constant angle $\psi$. 
\par
Then it is straightforward to obtain asymptotic relations
\begin{eqnarray} 
& & \sqrt{|1 - (z_1)^2| |1 - (z_2)^2|}  \nonumber \\ 
& = & 
\frac{1}{4} \sqrt{\left((r_1)^2 + (r_1)^{-2} - 2 \cos 2 \theta_1 \right) 
\left((r_2)^2 + (r_2)^{-2} - 2 \cos 2 \theta_2 \right)} \nonumber \\ 
& \sim& 
\frac{1}{4} \left(v^2 + v^{-2} - 2 \cos 2 \psi \right) ,  \ \ N \rightarrow \infty,
\end{eqnarray} 
\begin{eqnarray}
\label{asymp1}
& & v^{2 n} - v^{- 2n} - R^{2 n} + R^{- 2 n} \nonumber \\ 
& = & v^{2 n} - v^{- 2n} -v^{2 n}  \left( 1 - \frac{T}{N} \right)^{2 n} + 
v^{-2 n} \left( 1- \frac{T}{N} \right)^{- 2n} \nonumber \\ 
& \sim& v^{2 n} \left( 1 - e^{- 2 c T} \right),  \ \   N \rightarrow \infty, 
\end{eqnarray}
\and
\begin{eqnarray}
& & \left((\omega_1)^n + (\omega_1)^{-n} \right)  \left( ({\bar \omega_2})^n + ({\bar \omega_2})^{-n} \right)
\nonumber  \\
& = & \left( v^n \left( 1 - \frac{t_1}{N} \right )^n e^{ i n \left( \psi + \frac{\phi_1}{N}  \right) } + 
v^{-n} \left( 1 - \frac{t_1}{N} \right)^{- n} e^{ - i n \left( \psi + \frac{\phi_1}{N} \right) } 
\right) \nonumber \\ 
& & \times
\left( v^n \left( 1 - \frac{t_2}{N} \right )^n e^{- i n \left( \psi + \frac{\phi_2}{N} \right)} + 
v^{-n} \left( 1 - \frac{t_2}{N} \right)^{- n} e^{ i n\left( \psi + \frac{\phi_2}{N} \right)} \right) \nonumber \\
& \sim & v^{2 n} e^{-c (t_1 + t_2)} e^{i c (\phi_1 - \phi_2)}, \ \ N \rightarrow \infty
\end{eqnarray}
with $c = n/N$ fixed. Putting these asymptotic relations into (\ref{k1}), we find 
\begin{eqnarray}
\label{k1limit}
& & K_N^{(1)}(z_1,z_2) \nonumber \\ & \sim & \frac{4 N^2}{\pi} \frac{1}{v^2 + v^{-2} - 2 \cos 2 \psi} \int_0^1 dc 
\ c \ \frac{e^{-c (t_1 + t_2)} e^{i c (\phi_1 - \phi_2)}}{1 - e^{- 2 c T} }, \ \ N \rightarrow \infty. \nonumber \\ 
\end{eqnarray}
\par
Let us then examine the thin annulus limit $T \rightarrow 0$, which corresponds to a  strictly 
1D model on an ellipse. The effect of the radial asymmetry can readily be seen in that limit. Noting that $t_1,t_2 \rightarrow 0$ limit must also be taken because of the inequality $0 < t_1,t_2 < T$,  one obtains
\begin{eqnarray}
K_N^{(1)}(z_1,z_2) & \sim & \frac{2 N^2}{\pi T} \frac{1}{v^2 + v^{-2} - 2 \cos 2 \psi} 
\int_0^1 dc \ e^{i c (\phi_1 - \phi_2)} \nonumber \\ 
& = & \rho^{(1)}_v(\psi) \ e^{i (\phi_1 - \phi_2)/2} \ \frac{\sin\left\{ (\phi_1 - \phi_2)/2 \right\} }{ (\phi_1 - \phi_2)/2 }, 
\end{eqnarray}
where $\rho^{(1)}_v(\psi)$ is the asymptotic molecule density
\begin{equation}
\rho^{(1)}_v(\psi) = \frac{2 N^2}{\pi T} \frac{1}{v^2 + v^{-2} - 2 \cos 2 \psi} 
\end{equation}
on the edge of the elliptic region ${\cal E}$. Let us normalize the asymptotic molecule density as
\begin{equation}
\sigma(\psi) =  \frac{(v^2 - v^{-2})T}{4 N^2} \rho^{(1)}_v(\psi) =  \frac{v^2 - v^{-2}}{2 \pi}\frac{1}{v^2 + v^{-2} - 2 \cos 2 \psi} 
\end{equation}
and depict it in Figure 1. In Figure 1 we can see a transition from a strong radial asymmetry case ($v=1.1$)  to a weak radial asymmetry case ($v=2$). The weak radial asymmetry limit $v \rightarrow \infty$ results in a uniform density. As the phase factor $e^{i (\phi_j - \phi_\ell)/2} $ can be replaced with $1$ without changing the determinant forms (\ref{rhodet}) of the $k$-molecule correlation functions $\rho(z_1,z_2,\cdots,z_k)$, we arrive at
\begin{equation}
\label{sine-kernel-on-circle}
\frac{\rho(z_1,z_2,\cdots,z_k)}{\left\{ \rho^{(1)}_v(\psi) \right\}^k} 
 \sim \det\left[ \frac{\sin\left\{ (\phi_j - \phi_{\ell})/2 \right\} }{ (\phi_j - \phi_{\ell})/2 } 
\right]_{j,\ell= 1,2,\cdots,k}.
\end{equation}
This formula gives the elliptic version of the sine kernel formula for the 1D log-gas on a circle or the eigenvalues of the circular unitary ensemble (CUE) of random matrices\cite{mehta}. 

\begin{figure}[ht]
\epsfxsize=14cm
\centerline{\epsfbox{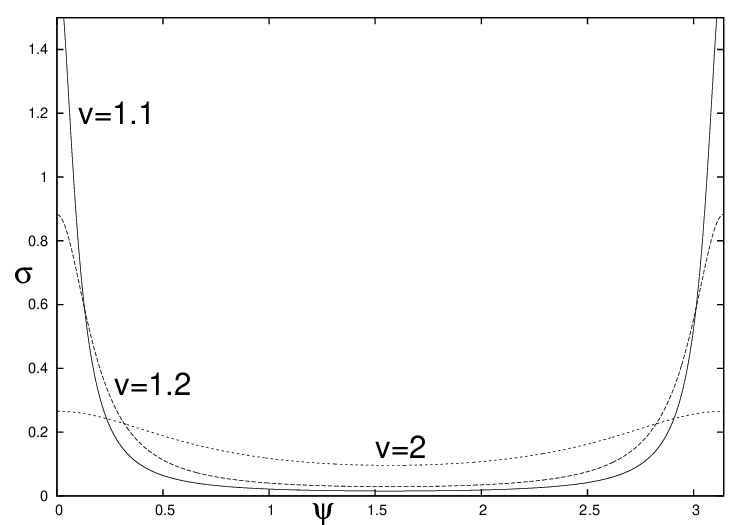}}
\caption{The normalized asymptotic molecule density  $\sigma(\psi)$.} 
\end{figure}

As discussed above, in a thin annulus limit $T \rightarrow 0$, the molecule 
correlation functions are given by the sine kernel formula (\ref{sine-kernel-on-circle}), which is well-known. Therefore, in order to see the transition from a quasi 1D model to a strictly 1D model, we next need to consider the opposite thick annulus limit $T \rightarrow \infty$. In the limit $T \rightarrow \infty$ of (\ref{k1limit}), we introduce a normalized 
two-molecule correlation function as
\begin{equation}
\lambda(\varphi) = 1 - \frac{\kappa(\tau,\varphi) \kappa(\tau,-\varphi)}{\kappa(\tau,0)^2},
\end{equation}
where
\begin{equation}
\kappa(\tau,\varphi) = \lim_{T \rightarrow \infty} 
\int_0^1 dc \ c \ \frac{e^{-c (t_1 + t_2)} e^{i c (\phi_1 - \phi_2)}}{1 - e^{- 2 c T} } = 
\frac{(- \tau + i \varphi - 1) e^{-\tau + i \varphi} + 1}{(-\tau + i \varphi)^2}
\end{equation}
with $\tau = t_1 + t_2$ and $\varphi= \phi_1 - \phi_2$. The normalized two-molecule 
correlation function $\lambda(\varphi)$ is depicted in Figure 2.

\begin{figure}[ht]
\epsfxsize=14cm
\centerline{\epsfbox{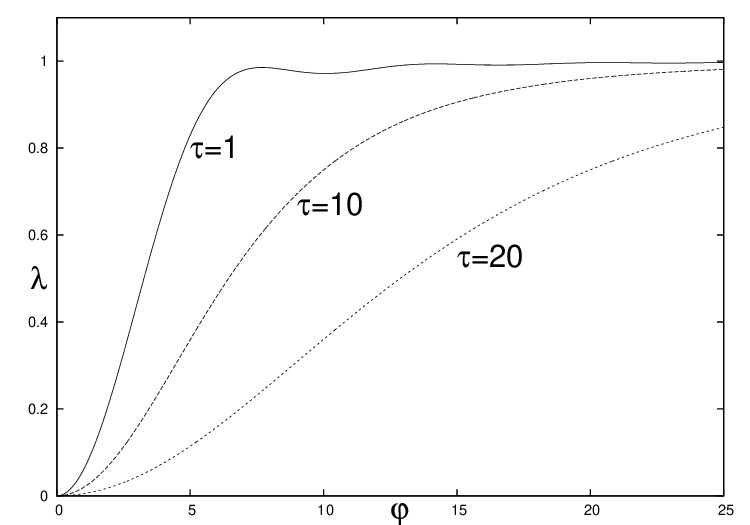}}
\caption{The normalized two-molecule correlation function $\lambda(\varphi)$.} 
\end{figure}

In another quasi 1D case, when the elliptic annulus is in the neighborhood of the real 
interval $[-1,1]$,  it is useful to modify the definitions of the scaling parameters 
$t_1$, $t_2$, $\phi_1$, $\phi_2$ and $T$ as
\begin{equation}
\label{scaling4}
r_1 = 1 +  \frac{t_1}{N}, \ \ \ r_2 = 1 + \frac{t_2}{N}, 
\end{equation}
\begin{equation}
\label{scaling5}
\theta_1 = \psi + \frac{\phi_1}{N}, \ \ \  \theta_2 = \psi + \frac{\phi_2}{N}
\end{equation}
and
\begin{equation}
\label{scaling6}
R = 1 + \frac{T}{N}.
\end{equation}
We additionally define a parameter $u$ as
\begin{equation}
\label{scaling7}
v = 1 +\frac{u}{N}
\end{equation}
and suppose that $0 < T < t_1,t_2< u$. The constant angle $\psi$ is 
in the region $0 \leq  \psi < 2 \pi$. When the elliptic annulus is in the neighborhood of  $[-1,1]$, 
only the cases $\psi=0$, $\psi=\pi/2$ and $\psi = \pi$ are considered. 
\par
As before we can  derive asymptotic relations
\begin{eqnarray} 
& & \sqrt{|1 - (z_1)^2| |1 - (z_2)^2|}  \nonumber \\ 
& = & 
\frac{1}{4} \sqrt{\left\{ \left(1 + \frac{t_1}{N} \right)^2 + \left(1 + \frac{t_1}{N} \right)^{-2} 
- 2 \cos 2 \left( \psi + \frac{\phi_1}{N} \right)  \right\} }
\nonumber \\  
& & \times 
\sqrt{ \left\{ \left(1 + \frac{t_2}{N} \right)^2 + \left(1 + \frac{t_2}{N} \right)^{-2} 
- 2 \cos 2 \left(\psi + \frac{\phi_2}{N}  \right) \right\} } \nonumber \\ 
& \sim& \left\{  \begin{array}{ll}  |s_1| |s_2|/N^2,  &  \psi = 0 \ {\rm or} \ \psi = \pi, \\ 
 \left(1  -  \cos 2 \psi \right)/2 ,  & \psi \neq 0 \ {\rm and} \ \psi \neq \pi,  \end{array} \right. \ \ N \rightarrow \infty,
\end{eqnarray} 
\begin{equation}
\label{asymp2}
v^{2 n} - v^{- 2n} - R^{2 n} + R^{- 2 n} \sim e^{2 c u} - e^{- 2 c u} - e^{ 2 c T} + e^{- 2 c T},  \ \   N \rightarrow \infty
\end{equation}
and
\begin{eqnarray}
& & \left((\omega_1)^n + (\omega_1)^{-n} \right)  \left( ({\bar \omega_2})^n + ({\bar \omega_2})^{-n} \right)
\nonumber  \\
& \sim & (e^{c s_1} e^{i n \psi} + e^{-c s_1} e^{ -  i n \psi} )
(e^{ c {\bar s_2}} e^{-i n \psi} + e^{- c {\bar s_2}} e^{i n \psi}), \ \ N \rightarrow \infty,
\end{eqnarray}
where 
\begin{equation}
s_j = t_j+ i \phi_j, \ \ j=1,2
\end{equation}
and $c = n/N$ fixed. 
\par
In the case $\psi = 0$ or $\psi = \pi$, these asymptotic relations 
and  (\ref{k1}) yield
\begin{equation} 
\label{k1limit-psi=0}
K^{(1)}(z_1,z_2) \sim 
\frac{N^4}{\pi} \frac{1}{|s_1| |s_2|}  
\int_0^1 dc \  c \ 
\frac{(e^{c s_1} + e^{-c s_1} ) (
e^{c {\bar s_2}} + e^{-c {\bar s_2}} )}{
e^{2 c u} - e^{-2 c u}  - e^{2 c T} + e^{-2 c T}} 
\end{equation}
in the limit $N \rightarrow \infty$. Let us consider the thin annulus limit shrinking 
to the interval $[-1,1]$. We first put $T = 0$ 
and then take the limit $u \rightarrow 0$. Keeping in mind that $u \rightarrow 0$ 
implies $t_1,t_2 \rightarrow 0$, we obtain
\begin{eqnarray}
\label{k1limit-sine}
K_N^{(1)}(z_1,z_2) 
& \sim & \frac{N^4}{4 \pi u} \frac{1}{|\phi_1| |\phi_2|} \int_0^1 dc \ 
(e^{i c \phi_1} + e^{- i c \phi_1})  
(e^{i c \phi_2} + e^{- i c \phi_2})  \nonumber \\
& = & \frac{N^4}{2 \pi u} \frac{1}{|\phi_1 \phi_2|} \left( 
\frac{\sin(\phi_1 - \phi_2)}{\phi_1 - \phi_2} + \frac{\sin(\phi_1 + \phi_2)}{\phi_1 + \phi_2} 
\right).
\end{eqnarray}
\par
When $\displaystyle \psi = \frac{\pi}{2}$,  noting $e^{2 i n \psi} = (-1)^n$ and evaluating the 
sums in (\ref{k1}) for odd $n$ and even $n$ separately, we find
\begin{eqnarray}
\label{k1limit-psi=pi/2}
& & K_N^{(1)}(z_1,z_2) =  \frac{1}{\displaystyle \sqrt{|1 - (z_1)^2| |1 - (z_2)^2|}}  \nonumber \\ 
& & \times \left\{  \frac{1}{2 \pi \log(v/R)} + \frac{1}{\pi}  \sum_{n=1}^{N-1} n 
\frac{((\omega_1)^n + (\omega_1)^{-n}) ( ({\bar \omega_2})^n + ({\bar \omega_2})^{-n})}{
v^{2 n} - v^{- 2n} - R^{2 n} + R^{- 2 n}}
\right\} \nonumber \\ 
& = &  \frac{1}{\displaystyle \sqrt{|1 - (z_1)^2| |1 - (z_2)^2|}}   \Biggl\{  \frac{1}{2 \pi \log(v/R)}  
\nonumber \\ 
& & + \frac{1}{\pi}  \sum_{n=0}^{[N/2 - 1]} (2 n + 1)
\frac{((\omega_1)^{2 n + 1}+ (\omega_1)^{-2 n - 1}) ( ({\bar \omega_2})^{2 n + 1} + ({\bar \omega_2})^{-2 n - 1})}{
v^{4 n + 2} - v^{- 4 n - 2} - R^{4 n +2} + R^{- 4 n - 2}} \nonumber \\ 
& & +  \frac{1}{\pi}  \sum_{n=1}^{[(N-1)/2]} (2 n) 
\frac{((\omega_1)^{2 n} + (\omega_1)^{-2 n}) ( ({\bar \omega_2})^{2 n} + ({\bar \omega_2})^{-2 n})}{
v^{4 n} - v^{- 4n} - R^{4 n} + R^{- 4 n}} 
\Biggl\} \nonumber \\ 
& \sim &  
\frac{N^2}{2 \pi} \int_0^1 dc \  c \ 
\frac{  e^{c (s_1 + {\bar s_2})} + e^{-c (s_1+ {\bar s_2})}   -  e^{c (s_1 - {\bar s_2})} -  e^{-c (s_1- {\bar s_2})}  }{
e^{2 c u} - e^{-2 c u}  - e^{2 c T} + e^{-2 c T}} 
\nonumber \\ 
&  &  +  
\frac{N^2}{2 \pi} \int_0^1 dc \  c \ 
\frac{  e^{c (s_1 + {\bar s_2})} + e^{-c (s_1+ {\bar s_2})}    + e^{c (s_1 - {\bar s_2})} + e^{-c (s_1- {\bar s_2})}  }{
e^{2 c u} - e^{-2 c u}  - e^{2 c T} + e^{-2 c T}}  \nonumber \\ 
& = &  
\frac{N^2}{\pi} \int_0^1 dc \  c \ 
\frac{e^{c (s_1 + {\bar s_2})} + e^{-c (s_1+ {\bar s_2})} }{
e^{2 c u} - e^{-2 c u}  - e^{2 c T} + e^{-2 c T}} 
\end{eqnarray}
in the limit $N \rightarrow \infty$. Here $[x]$ is the largest integer less than or equal to $x$. Taking the thin annulus limit $u \rightarrow 0$ after putting $T=0$ 
results in
\begin{equation}
\label{sine-kernel}
K_N^{(1)}(z_1,z_2)  \sim 
\frac{N^2}{2 \pi u}  \int_0^1 dc \ \cos\left( c (\phi_1 - \phi_2) \right) = 
\rho^{(1)}_u \frac{\sin(\phi_1 - \phi_2)}{\phi_1 - \phi_2}, 
\end{equation}
where
$
\displaystyle 
\rho^{(1)}_u=\frac{N^2}{2 \pi u} .
$
The $k$-molecule correlation functions $\rho(z_1,z_2,\cdots,z_k)$ 
accordingly satisfy
\begin{equation}
\frac{\rho(z_1,z_2,\cdots,z_k)}{\left\{ \rho^{(1)}_u \right\}^k} 
 \sim \det\left[ \frac{\sin (\phi_j - \phi_{\ell}) }{ \phi_j - \phi_{\ell} } 
\right]_{j,l = 1,2,\cdots,k}.
\end{equation}
This formula reproduces the sine kernel formula for the log-gas on a 1D line originally known from  the theory of the Gaussian unitary ensemble (GUE) of random matrices\cite{mehta}. 

\subsection{Model II}

Next we consider Model II with the weight function (\ref{wa2}) and 
the kernel function $K_N^{(2)}(z_1,z_2)$ in (\ref{k2}). The polar 
coordinates $r_1,r_2 ,\theta_1$ and $\theta_2$ are 
introduced in (\ref{polar}) as before. 
\par
If one is interested in the 
elliptic annulus in  the neighborhood 
of the outer edge $r_1,r_2 = v > 1$, the scaling parameters 
$t_1$, $t_2$, $\phi_1$, $\phi_2$ and $T$ defined in (\ref{scaling1}), 
(\ref{scaling2}) and (\ref{scaling3}) are again useful. 
Let us first derive asymptotic relations
\begin{eqnarray}
& & (\omega_1 - \omega_1^{-1}) ({\bar \omega_2} - {\bar \omega_2}^{-1}) 
\nonumber \\ 
& = &  \left( v  \left( 1 - \frac{t_1}{N} \right ) e^{ i \left( \psi + \frac{\phi_1}{N}  \right) } -
v^{-1} \left( 1- \frac{t_1}{N} \right)^{- 1} e^{ - i \left( \psi + \frac{\phi_1}{N} \right) } 
\right) \nonumber \\ 
& & \times
\left( v \left( 1 - \frac{t_2}{N} \right ) e^{- i \left( \psi + \frac{\phi_2}{N} \right)} -
v^{-1} \left( 1 - \frac{t_2}{N} \right)^{- 1} e^{ i \left( \psi + \frac{\phi_2}{N} \right)} \right) \nonumber \\
& \sim & v^2 + v^{-2} - 2 \cos\psi, \ \ N \rightarrow \infty
\end{eqnarray}
and
\begin{eqnarray}
& & \left((\omega_1)^n - (\omega_1)^{-n} \right)  \left( ({\bar \omega_2})^n - ({\bar \omega_2})^{-n} \right)
\nonumber  \\
& = & \left( v^{n} \left( 1 - \frac{t_1}{N} \right )^n e^{ i n \left( \psi + \frac{\phi_1}{N}  \right) } -
v^{-n} \left( 1 - \frac{t_1}{N} \right)^{- n} e^{ - i n \left( \psi + \frac{\phi_1}{N} \right) } 
\right) \nonumber \\ 
& & \times
\left( v^n \left( 1 - \frac{t_2}{N} \right )^n e^{- i n \left( \psi + \frac{\phi_2}{N} \right)} -
v^{-n} \left( 1 - \frac{t_2}{N} \right)^{- n} e^{ i n\left( \psi + \frac{\phi_2}{N} \right)} \right) \nonumber \\
& \sim & v^{2 n} e^{-c (t_1 + t_2)} e^{i c (\phi_1 - \phi_2)}, \ \ N \rightarrow \infty
\end{eqnarray}
with $c = n/N$ fixed. We put these asymptotic relations and (\ref{asymp1}) into (\ref{k2}) 
and arrive at
\begin{eqnarray}
\label{k2limit}
& & K_N^{(2)}(z_1,z_2) \nonumber \\ & \sim & \frac{4 N^2}{\pi} \frac{1}{v^2 + v^{-2} - 2 \cos 2 \psi} \int_0^1 dc 
\ c \ \frac{e^{-c (t_1 + t_2)} e^{i c (\phi_1 - \phi_2)}}{1 - e^{- 2 c T} }, \ \ N \rightarrow \infty. \nonumber \\ 
\end{eqnarray}
This is exactly equal to the asymptotic result in (\ref{k1limit}) for Model I. We thus expect 
it to be a universal asymptotic behavior. As before it yields the elliptic version of 
the sine kernel formula in the thin annulus limit $T \rightarrow 0$. 
\par
We then consider an elliptic annulus in the neighborhood of the real interval $[-1,1]$. 
The scaling parameters $t_1$, $t_2$, $\phi_1$, $\phi_2$ and $T$ are modified and an 
additional parameter $u$ is introduced according to  (\ref{scaling4}), 
(\ref{scaling5}), (\ref{scaling6}) and (\ref{scaling7}). Resulting asymptotic 
relations are
\begin{eqnarray}
& & (\omega_1 - \omega_1^{-1}) ({\bar \omega_2} - {\bar \omega_2}^{-1}) 
\nonumber \\ 
& = & \left( \left( 1 + \frac{t_1}{N} \right ) e^{ i \left( \psi + \frac{\phi_1}{N}  \right) } -
\left( 1 + \frac{t_1}{N} \right)^{- 1} e^{ - i \left( \psi + \frac{\phi_1}{N} \right) } 
\right) \nonumber \\ 
& & \times
\left( \left( 1 + \frac{t_2}{N} \right ) e^{- i \left( \psi + \frac{\phi_2}{N} \right)} -
\left( 1 +  \frac{t_2}{N} \right)^{- 1} e^{ i \left( \psi + \frac{\phi_2}{N} \right)} \right) \nonumber \\
& \sim& \left\{  \begin{array}{ll}  4 s_1 {\bar s_2} /N^2,  &  
\psi = 0 \ {\rm or} \ \psi = \pi, \\ 
 2   -  2 \cos 2 \psi,  & \psi \neq 0 \ {\rm and} \ \psi \neq \pi,  \end{array} \right. \ \ N \rightarrow \infty
\end{eqnarray}
and
\begin{eqnarray}
& & \left((\omega_1)^n - (\omega_1)^{-n} \right)  \left( ({\bar \omega_2})^n - ({\bar \omega_2})^{-n} \right)
\nonumber  \\
& \sim & (e^{c s_1} e^{i n \psi} - e^{-c s_1} e^{ -  i n \psi} )
(e^{ c {\bar s_2}} e^{-i n \psi} - e^{- c {\bar s_2}} e^{i n \psi}), \ \ N \rightarrow \infty,
\end{eqnarray}
where $s_j = t_j+ i \phi_j, \ \ j=1,2$ and $c = n/N$ fixed. 
\par
In the case $\psi = 0$ or $\psi = \pi$, it follows from these asymptotic relations, (\ref{k2}) and (\ref{asymp2}) that
\begin{equation} 
\label{k2limit-psi=0}
K_N^{(2)}(z_1,z_2) \sim 
\frac{N^4}{\pi} \frac{1}{s_1 {\bar s_2}}  
\int_0^1 dc \  c \ 
\frac{(e^{c s_1} - e^{-c s_1} ) (
e^{c {\bar s_2}} - e^{-c {\bar s_2}} )}{
e^{2 c u} - e^{-2 c u}  - e^{2 c T} + e^{-2 c T}} 
\end{equation}
in the limit $N \rightarrow \infty$. The difference between the asymptotic limit (\ref{k1limit-psi=0}) 
of Model I correlation functions and the corresponding limit (\ref{k2limit-psi=0}) of Model II 
reveals  a breakdown of the universality around the edges of the 
interval $[-1,1]$. In order to examine the thin annulus limit, let us put $T = 0$ and then take 
the limit $u \rightarrow 0$. We obtain
\begin{eqnarray}
\label{k2limit-sine}
K_N^{(2)}(z_1,z_2) 
& \sim & \frac{N^4}{4 \pi u} \frac{1}{\phi_1\phi_2} \int_0^1 dc \ 
(e^{i c \phi_1} - e^{- i c \phi_1})  
(e^{i c \phi_2} - e^{- i c \phi_2})  \nonumber \\
& = & \frac{N^4}{2 \pi u} \frac{1}{\phi_1 \phi_2} \left( 
\frac{\sin(\phi_1 - \phi_2)}{\phi_1 - \phi_2} - \frac{\sin(\phi_1 + \phi_2)}{\phi_1 + \phi_2} 
\right),
\end{eqnarray}
which shows the striking sign difference of  Model  I and Model II results.
\par
When $\displaystyle \psi = \frac{\pi}{2}$,  as before we can derive
\begin{equation} 
\label{k2limit-psi=pi/2}
K_N^{(2)}(z_1,z_2) \sim 
\frac{N^2}{\pi} \int_0^1 dc \  c \ 
\frac{e^{c (s_1 + {\bar s_2})} + e^{-c (s_1+ {\bar s_2})} }{
e^{2 c u} - e^{-2 c u}  - e^{2 c T} + e^{-2 c T}} 
\end{equation}
in the limit $N \rightarrow \infty$. This is exactly equal to the corresponding limit (\ref{k1limit-psi=pi/2}) 
of Model I result. Thus the universal behavior is recovered.

\subsection{Model III}

Model III has the weight function (\ref{wa3}) and 
the kernel function $K_N^{(3)}(z_1,z_2)$ in (\ref{k3}). 
We use the polar coordinates $r_1,r_2 ,\theta_1$ and $\theta_2$ 
introduced in (\ref{polar}). 
\par
In order to study the elliptic annulus in  the neighborhood 
of the outer edge $r_1,r_2 = v > 1$,  we again employ 
the scaling parameters $t_1$, $t_2$, $\phi_1$, $\phi_2$ and $T$ defined in (\ref{scaling1}), 
(\ref{scaling2}) and (\ref{scaling3}) . Then we find asymptotic relations
\begin{eqnarray} 
| 1 - z_j | & = & \left| 1 - \frac{1}{2} \left( \omega_j + \frac{1}{\omega_j} \right) \right| = 
\left| 1 - \frac{1}{2} \left( r_j e^{i \theta_j} + \frac{1}{r_j} e^{- i \theta_j} \right) \right| 
 \nonumber \\ 
& = & 
\frac{1}{2} \left| \left( \sqrt{r_j} e^{i \theta_j/2} - \frac{1}{\sqrt{r_j}} e^{- i \theta_j/2} \right)^2 \right|
= \frac{1}{2} \left| \sqrt{r_j} e^{i \theta_j/2} - \frac{1}{\sqrt{r_j}} e^{- i \theta_j/2}  \right|^2
\nonumber \\ 
& = & \frac{1}{2} ( r_j + r_j^{-1} - 2 \cos\theta_j) \nonumber \\
& \sim & \frac{1}{2} ( v + v^{-1}- 2 \cos\psi) , \  \ \ N \rightarrow \infty
\end{eqnarray}
with $j = 1,2$,
\begin{eqnarray}
& & (\omega_1 + 1) ({\bar \omega_2} + 1)  =   (r_1 e^{i \theta_1} + 1) ( r_2 e^{- i \theta_2} + 1)
\nonumber \\ & \sim & (v e^{i \psi} + 1) ( v e^{-i\psi} + 1) = v ( v + v^{-1} + 2 \cos\psi), \  \ N \rightarrow \infty
\end{eqnarray}
and
\begin{eqnarray}
& &  \left((\omega_1)^{n+1} + (\omega_1)^{-n} \right)  \left( ({\bar \omega_2})^{n+1} + ({\bar \omega_2})^{-n} \right) 
 \nonumber \\ & \sim & v^{2 (n+1)} e^{-c (t_1 + t_2)} e^{i c (\phi_1 - \phi_2)}, \ \ N \rightarrow \infty
\end{eqnarray}
with $c = n/N$ fixed. Putting these asymptotic relations and
\begin{equation}
\label{asymp3}
v^{2 n + 1} - v^{- 2n- 1} - R^{2 n + 1} + R^{- 2 n - 1} \sim v^{2 n+ 1} \left( 1 - e^{- 2 c T} \right),  \ \   N \rightarrow \infty 
\end{equation}
into  (\ref{k3}) yields
\begin{eqnarray}
\label{k3limit}
& & K_N^{(3)}(z_1,z_2) \nonumber \\ & \sim & \frac{4 N^2}{\pi} \frac{1}{v^2 + v^{-2} - 2 \cos 2 \psi} \int_0^1 dc 
\ c \ \frac{e^{-c (t_1 + t_2)} e^{i c (\phi_1 - \phi_2)}}{1 - e^{- 2 c T} }, \ \ N \rightarrow \infty. \nonumber \\ 
\end{eqnarray}
This should be a universal asymptotic behavior, because it is again identical  to the results in (\ref{k1limit})
and (\ref{k2limit})  for Model I and Model II. 
\par
Let us modify the scaling parameters $t_1$, $t_2$, $\phi_1$, $\phi_2$ and $T$,  and introduce an 
additional parameter $u$ as written in (\ref{scaling4}), (\ref{scaling5}), (\ref{scaling6}) and (\ref{scaling7}). 
Then, in the limit $N \rightarrow \infty$, the  elliptic annulus shrinks into the neighborhood of 
the real interval $[-1,1]$. We require asymptotic formulas
\begin{eqnarray}
| 1 - z_j | & = & \frac{1}{2} (r_j + r_j^{-1} - 2 \cos\theta_j ) \nonumber \\ 
& = & \frac{1}{2}  \left( \left( 1 + \frac{t_j}{N} \right ) + 
\left( 1 + \frac{t_j}{N} \right)^{- 1} - 2 \cos\left( \psi + \frac{\phi_j}{N}  \right)
\right) \nonumber \\ 
& \sim& \left\{  \begin{array}{ll}  |s_j|^2 /(2 N^2),  &  
\psi = 0,  \\ 
(2   -  2 \cos \psi)/2,  & \psi \neq 0,  \end{array} \right. \ \ N \rightarrow \infty, 
\end{eqnarray}
\begin{eqnarray}
\omega_j + 1  & = & r_j e^{i \theta_j} + 1 \nonumber \\ 
& = & \left( 1 + \frac{t_j}{N} \right ) e^{i ( \psi + \phi_j/N)} + 1
 \nonumber \\ 
& \sim& \left\{  \begin{array}{ll}  - s_j/N&  
\psi = \pi,  \\ 
e^{i \psi} + 1,  & \psi \neq \pi,  \end{array} \right. \ \ N \rightarrow \infty
\end{eqnarray}
with $j = 1,2$, 
\begin{eqnarray}
& & \left((\omega_1)^{n+1} + (\omega_1)^{-n} \right)  \left( ({\bar \omega_2})^{n+1} + ({\bar \omega_2})^{-n} \right)
\nonumber  \\
& \sim & (e^{c s_1} e^{i (n+1) \psi} + e^{-c s_1} e^{ -  i n \psi} )
(e^{ c {\bar s_2}} e^{-i (n+1) \psi} +  e^{- c {\bar s_2}} e^{i n \psi}), \nonumber \\ 
& & N \rightarrow \infty
\end{eqnarray}
and
\begin{equation}
\label{asymp4}
v^{2 n+1} - v^{- 2n-1} - R^{2 n+1} + R^{- 2 n-1} \sim e^{2 c u} - e^{- 2 c u} - e^{ 2 c T} + e^{- 2 c T},  \ \   N \rightarrow \infty.
\end{equation}
Here $s_j = t_j+ i \phi_j, \ \ j=1,2$ and $c = n/N$ fixed. 
\par
In the case $\psi = 0$,  these asymptotic relations and (\ref{k3}) lead to 
\begin{equation} 
K_N^{(3)}(z_1,z_2) \sim 
\frac{N^4}{\pi} \frac{1}{|s_1| |s_2| }  
\int_0^1 dc \  c \ 
\frac{(e^{c s_1} + e^{-c s_1} ) (
e^{c {\bar s_2}} + e^{-c {\bar s_2}} )}{
e^{2 c u} - e^{-2 c u}  - e^{2 c T} + e^{-2 c T}} 
\end{equation}
in the limit $N \rightarrow \infty$. This is equal to the 
asymptotic limit (\ref{k1limit-psi=0}) of $K_N^{(1)}(z_1,z_2)$ with $\psi = 0$. 
On the other hand, in the case $\psi = \pi$, we obtain
\begin{equation} 
K_N^{(3)}(z_1,z_2) \sim 
\frac{N^4}{\pi} \frac{1}{s_1 {\bar s_2}}  
\int_0^1 dc \  c \ 
\frac{(e^{c s_1} - e^{-c s_1} ) (
e^{c {\bar s_2}} - e^{-c {\bar s_2}} )}{
e^{2 c u} - e^{-2 c u}  - e^{2 c T} + e^{-2 c T}} 
\end{equation}
in the limit $N \rightarrow \infty$. This is equal to the asymptotic limit (\ref{k2limit-psi=0}) of 
$K_N^{(2)}(z_1,z_2)$ with $\psi = \pi$.  Around the right edge of the interval $[-1,1]$, Model III 
behaves like Model I, and around the left edge, it behaves like Model  II. This is expected from 
the forms of the weight functions.
\par
Putting $\displaystyle \psi = \frac{\pi}{2}$,  we find
\begin{equation} 
\label{k3limit-psi=pi/2}
K_N^{(3)}(z_1,z_2) \sim 
\frac{N^2}{\pi} \int_0^1 dc \  c \ 
\frac{e^{c (s_1 + {\bar s_2})} + e^{-c (s_1+ {\bar s_2})} }{
e^{2 c u} - e^{-2 c u}  - e^{2 c T} + e^{-2 c T}} 
\end{equation}
in the limit $N \rightarrow \infty$. This is again equal to the corresponding limits (\ref{k1limit-psi=pi/2}) 
and (\ref{k2limit-psi=pi/2}). It is  the universal behavior expected when $\displaystyle \psi = \pi/2$.

\subsection{Model IV}

We finally consider Model IV. The weight function is (\ref{wa4}) and 
the kernel function $K_N^{(4)}(z_1,z_2)$ is (\ref{k4}).  Let us as before 
use the polar coordinates $r_1,r_2 ,\theta_1$ and $\theta_2$ 
introduced in (\ref{polar}). 
\par
The scaling parameters $t_1$, $t_2$, $\phi_1$, $\phi_2$ and $T$ in (\ref{scaling1}), 
(\ref{scaling2}) and (\ref{scaling3})  are again used in order to magnify the region 
in  the neighborhood of the outer edge $r_1,r_2 = v > 1$.  The asymptotic relations
\begin{equation} 
| 1 + z_j | = \frac{1}{2} ( r_j + r_j^{-1} + 2 \cos\theta_j) \sim \frac{1}{2} ( v + v^{-1} +  2 \cos\psi) , \  \ \ N \rightarrow \infty
\end{equation}
with $j = 1,2$,
\begin{equation}
(\omega_1 -  1) ({\bar \omega_2} -  1)  =   (r_1 e^{i \theta_1} - 1) ( r_2 e^{- i \theta_2} - 1) \sim v ( v + v^{-1} -  2 \cos\psi), \  \ N \rightarrow \infty
\end{equation}
and
\begin{eqnarray}
& &  \left((\omega_1)^{n+1} - (\omega_1)^{-n} \right)  \left( ({\bar \omega_2})^{n+1} - ({\bar \omega_2})^{-n} \right) 
\nonumber  \\ &  \sim & v^{2 (n+1)} e^{-c (t_1 + t_2)} e^{i c (\phi_1 - \phi_2)}, \ \ N \rightarrow \infty
\end{eqnarray}
follow with $c = n/N$ fixed. One puts these relations and (\ref{asymp3}) into  (\ref{k4}). Then one obtains
\begin{eqnarray}
\label{k4limit}
& & K_N^{(4)}(z_1,z_2) \nonumber \\ & \sim & \frac{4 N^2}{\pi} \frac{1}{v^2 + v^{-2} - 2 \cos 2 \psi} \int_0^1 dc 
\ c \ \frac{e^{-c (t_1 + t_2)} e^{i c (\phi_1 - \phi_2)}}{1 - e^{- 2 c T} }, \ \ N \rightarrow \infty. \nonumber \\ 
\end{eqnarray}
We see that it is again identified with the universal results in (\ref{k1limit}), (\ref{k2limit})  and ({\ref{k3limit}) 
for Model I, Model II and Model III, respectively. 
\par
On the other hand, in order to magnify the region in the neighborhood of the real interval $[-1,1]$, 
we redefine  the scaling parameters $t_1$, $t_2$, $\phi_1$, $\phi_2$, $T$ and $u$ as written in 
(\ref{scaling4}), (\ref{scaling5}), (\ref{scaling6}) and (\ref{scaling7}). 
Then, in the limit $N \rightarrow \infty$, asymptotic formulas
\begin{eqnarray}
| 1 + z_j | & = & \frac{1}{2} (r_j + r_j^{-1} + 2 \cos\theta_j ) \nonumber \\ 
& = & \frac{1}{2}  \left( \left( 1 + \frac{t_j}{N} \right ) + 
\left( 1 + \frac{t_j}{N} \right)^{- 1} + 2 \cos\left( \psi + \frac{\phi_j}{N}  \right)
\right) \nonumber \\ 
& \sim& \left\{  \begin{array}{ll}  |s_j|^2 /(2 N^2),  &  
\psi = \pi,  \\ 
(2   +  2 \cos \psi)/2,  & \psi \neq \pi ,  \end{array} \right. \ \ N \rightarrow \infty, 
\end{eqnarray}
\begin{eqnarray}
\omega_j - 1  & = & r_j e^{i \theta_j} - 1 \nonumber \\ 
& = & \left( 1 + \frac{t_j}{N} \right ) e^{i ( \psi + \phi_j/N)} - 1
 \nonumber \\ 
& \sim& \left\{  \begin{array}{ll}  s_j/N&  
\psi = 0,  \\ 
e^{i \psi} - 1,  & \psi \neq 0,  \end{array} \right. \ \ N \rightarrow \infty
\end{eqnarray}
with $j = 1,2$ and
\begin{eqnarray}
& & \left((\omega_1)^{n+1} - (\omega_1)^{-n} \right)  \left( ({\bar \omega_2})^{n+1} -  ({\bar \omega_2})^{-n} \right)
\nonumber  \\
& \sim & (e^{c s_1} e^{i (n+1) \psi} -  e^{-c s_1} e^{ -  i n \psi} )
(e^{ c {\bar s_2}} e^{-i (n+1) \psi} -   e^{- c {\bar s_2}} e^{i n \psi}), \nonumber \\ 
& & N \rightarrow \infty
\end{eqnarray}
are obtained. Here $s_j = t_j+ i \phi_j, \ \ j=1,2$ and $c = n/N$ fixed. 
\par
In the case $\psi = 0$,  from these asymptotic relations, (\ref{k4}) and (\ref{asymp4}), 
we find
\begin{equation} 
K_N^{(4)}(z_1,z_2) \sim 
\frac{N^4}{\pi} \frac{1}{s_1 {\bar s_2}}  
\int_0^1 dc \  c \ 
\frac{(e^{c s_1} - e^{-c s_1} ) (
e^{c {\bar s_2}} - e^{-c {\bar s_2}} )}{
e^{2 c u} - e^{-2 c u}  - e^{2 c T} + e^{-2 c T}} 
\end{equation}
in the limit $N \rightarrow \infty$. This is equal to the 
asymptotic limit (\ref{k2limit-psi=0}) of $K_N^{(2)}(z_1,z_2)$ with $\psi = 0$. 
In the case $\psi = \pi$, an asymptotic formula
\begin{equation} 
K_N^{(4)}(z_1,z_2) \sim 
\frac{N^4}{\pi} \frac{1}{|s_1| |s_2| }  
\int_0^1 dc \  c \ 
\frac{(e^{c s_1} + e^{-c s_1} ) (
e^{c {\bar s_2}} + e^{-c {\bar s_2}} )}{
e^{2 c u} - e^{-2 c u}  - e^{2 c T} + e^{-2 c T}} 
\end{equation}
is obtained in the limit $N \rightarrow \infty$. This is equal to the asymptotic 
limit (\ref{k1limit-psi=0}) of 
$K_N^{(1)}(z_1,z_2)$ with $\psi = \pi$. 
\par
In the remaining case $\displaystyle \psi = \frac{\pi}{2}$,  one arrives at
\begin{equation} 
K_N^{(4)}(z_1,z_2) \sim 
\frac{N^2}{\pi} \int_0^1 dc \  c \ 
\frac{e^{c (s_1 + {\bar s_2})} + e^{-c (s_1+ {\bar s_2})} }{
e^{2 c u} - e^{-2 c u}  - e^{2 c T} + e^{-2 c T}} 
\end{equation}
in the limit $N \rightarrow \infty$. This is equal to the universal limits (\ref{k1limit-psi=pi/2}), 
(\ref{k2limit-psi=pi/2}) and (\ref{k3limit-psi=pi/2}), as expected.

\section*{Summary and discussion}

2D Coulomb gases distributed on an elliptic annulus 
at a special inverse temperature $\beta = 2$ were studied,  and 
asymptotic forms of the correlation functions in the thermodynamic 
limit were evaluated. For an elliptic annulus contained in the neighborhood 
of the outer edge, we found universal asymptotic forms of the correlation 
functions.  When the annulus shrinks into the real interval $[-1,1]$, 
non-universal behavior was observed around the edges $-1$ 
and $1$, where the weight functions $w_{\cal A}(z)$ on the annulus 
have singular points.  The models considered in this paper are related 
to the Chebyshev polynomials, and their orthogonality on the complex 
plane were employed  in the analysis. More general  polynomials related 
to the Gegenbauer polynomials are also known to be orthogonal on the 
complex plane\cite{NAKP,ANPV}. There are also Coulomb gas models associated with such 
more general polynomials, and it is interesting to check the universal behavior 
by using such more general models. 

\section*{Acknowledgements}

The author acknowledges support by the Japan Society for the Promotion of Science (KAKENHI 20K03764). He also thanks Aron Wennman for pointing to the reference \cite{walsh2}.

\section*{Appendix A \ Radially symmetric case} 
\setcounter{equation}{0}
\renewcommand{\theequation}{A.\arabic{equation}}

In this Appendix, we study a conventional radially symmetric case with a 
probability density function 
\begin{equation}
P(z_1,z_2,\cdots,z_N) \propto \prod_{j=1}^N w_{\bf C}(z_j) \prod_{j<l}^N |z_j - z_l|^2
\end{equation}
with a one-molecule weight function 
\begin{equation}
w_{\bf C}(z) = \left\{ \begin{array}{ll}
w_{\cal D}(z), & z \in {\cal D}, \\
0, & z \notin {\cal D}, \end{array} \right.
\end{equation}
where ${\cal D}$ is an annulus
\begin{equation}
{\cal D} = \left\{ z \mid |z|^2  \leq (v/2)^2\ {\rm and} \ |z|^2  \geq (R/2)^2 \right\}
\end{equation}
on the complex plane. Here $1 < R < v$. Moreover the weight function is supposed to be
\begin{equation}
w_{\cal D}(z) = |z|^\gamma, 
\end{equation}
where $\gamma$ is an arbitrary real number.  Then $w_{\bf C}(z)$ is fully determined by the radius $|z|$. Such a radially symmetric case is conventional, because the corresponding 
monic orthogonal polynomials $M_n(z)$ are known to be monomials $z^n$. Here asymptotic limit of 
the kernel function is evaluated for comparison with the Coulomb gas on an elliptic annulus. 
The orthogonality constant $h_n$ for $M_n(z) = z^n$ is
\begin{equation}
h_n  = \frac{2 \pi}{2 n + \gamma + 2} \frac{v^{2 n + \gamma + 2} - R^{2 n + \gamma + 2}}{
 2^{2 n + \gamma + 2}},
\end{equation}
because of the integral
\begin{eqnarray}
 & & \int_{\bf C} \sqrt{w_{\bf  C}({\bar z}) w_{\bf C}(z)} M_m({\bar z}) M_n(z) dx dy \nonumber \\ 
 & = & \int_{R/2}^{v/2} dr \int_0^{2 \pi} d\theta \ r^{m + n + \gamma + 1}  e^{i (n - m) \theta} \nonumber \\ 
 & = & \delta_{mn} \frac{2 \pi}{2 n + \gamma + 2} \frac{v^{2 n + \gamma + 2} - R^{2 n + \gamma + 2}}{
 2^{2 n + \gamma + 2}}.
\end{eqnarray}
\par
Let us calculate the asymptotic limit of the kernel function
\begin{eqnarray}
& & K_{N,\gamma}^{(RS)}(z_1,z_2) 
= \sqrt{w_{\cal D}(z_1) w_{\cal D}({\bar z_2})} \sum_{n = 0}^{N-1} \frac{1}{h_n} M_n(z_1) M_n({\bar z_2}) 
\nonumber \\ 
& = & \frac{(r_1 r_2)^{\gamma/2}}{2 \pi} \sum_{n=0}^{N-1} (2 n + \gamma + 2) 
\frac{(r_1 r_2)^n e^{i n (\theta_1 - \theta_2)}}{
(v/2)^{2n + \gamma + 2} - (R/2)^{2 n + \gamma + 2}},
\end{eqnarray}
where $z_j = r_j e^{i \theta_j}$ with a positive $r_j$ and a real $\theta_j$ ($j = 1,2$). Introducing scaling parameters $t_1$, $t_2$, $\phi_1$, $\phi_2$ and $T$ as
\begin{equation}
r_1 = \frac{v}{2} \left( 1  - \frac{t_1}{N} \right) , \ \ \ r_2 = \frac{v}{2} \left( 1 - \frac{t_2}{N} \right),
\end{equation}
\begin{equation}
\theta_1 = \psi + \frac{\phi_1}{N}, \ \ \  \theta_2 = \psi + \frac{\phi_2}{N}
\end{equation}
and
\begin{equation}
R = \frac{v}{2} \left( 1 - \frac{T}{N} \right),
\end{equation}
where $0 < t_1,t_2< T$ and $0 \leq  \psi < 2 \pi$, we find
\begin{equation}
K_{N,\gamma}^{(RS)}(z_1,z_2) \sim \frac{4 N^2}{\pi v^2} \int_0^1 dc \ c \ 
\frac{e^{-c(t_1 + t_2)} e^{i c (\phi_1 - \phi_2)}}{
1 - e^{- 2 c T}}
\end{equation}
in the limit $N \rightarrow \infty$ with $c = n/N$ and $\gamma$ fixed.  This is equal 
to the $v \rightarrow \infty$ limit of the universal formula (\ref{k1limit}), (\ref{k2limit}), (\ref{k3limit}) and (\ref{k4limit}) 
as expected, and itself shows a universal behavior because it is independent of $\gamma$. 

\section*{Appendix B \ Bessel kernel} 
\setcounter{equation}{0}
\renewcommand{\theequation}{B.\arabic{equation}}

In this Appendix, we consider another conventional model with a probability density function 
\begin{equation}
P(x_1,x_2,\cdots,x_N) \propto \prod_{j=1}^N w_{\bf R}(x_j) \prod_{j<l}^N |x_j - x_l|^2.
\end{equation}
It is a 1D model and real numbers $x_1,x_2,\cdots,x_N$ are molecule locations. 
A one-molecule weight function is
\begin{equation}
w_{\bf R}(x) = \left\{ \begin{array}{ll}
w_{[-1,1]}(x),  & x \in [-1,1], \\
0, & x \notin [-1,1],  \end{array} \right.
\end{equation}
where $x$ is real and  $w_{[-1,1]}(x) = (1 - x)^a (1 + x)^b$ with $a,b > -1$. 
This model is called the Jacobi unitary ensemble 
in random matrix theory\cite{FK,NW,NS}. In the limit that the elliptic annulus shrinks into the interval $[-1,1]$, the 2D Coulomb gas model is reduced to this 1D model.
\par
The monic orthogonal polynomials corresponding to the Jacobi unitary ensemble is
\begin{equation}
M_n(x) = 2^n n! \frac{\Gamma(a + b + n + 1)}{\Gamma(a + b + 2 n + 1)} 
P_n^{(a,b)}(x)
\end{equation}
and the orthogonality constant $h_n$ is
\begin{equation}
h_n = 2^{2 n + a  + b + 1} n! \frac{\Gamma(a +  n + 1) \Gamma(b +  n + 1) \Gamma(a + b + n + 1)}{
\Gamma(a + b  + 2 n + 1) \Gamma(a + b + 2 n + 2)}.
\end{equation}
Here $\Gamma(x)$ is the Gamma function and $P_n^{(a,b)}(x)$ are the Jacobi polynomials
\begin{equation}
P_n^{(a,b)}(x) = \frac{(-1)^n}{2^n n!} (1 - x)^{-a} (1 + x)^{-b} 
\frac{d^n}{d x^n} \left\{ (1 -  x)^a (1 + x)^b (1 - x^2)^n \right\}.
\end{equation}
Using an asymptotic form of the Jacobi polynomials\cite{szego2}
\begin{equation}
P_n^{(a,b)}\left(1 - \frac{\phi^2}{2 N^2} \right) \sim \left(  \frac{\phi}{2 N} \right)^{-a} 
J_a(c \phi)
\end{equation}
in the limit $N \rightarrow \infty$ with $c = n/N$ and $\phi > 0$ fixed, we can derive the 
asymptotic form of the kernel function\cite{NS,pjf3,VZ,TW}
\begin{eqnarray}
\label{bessel-kernel}
& & K_{N,a,b}^{Jacobi}(x_1,x_2) =  \sqrt{w_{[-1,1]}(x_1) w_{[-1,1]}(x_2)}  \sum_{n=0}^{N-1} 
\frac{1}{h_n} M_n(x_1) M_n(x_2) \nonumber \\ 
& & \sim K_a^{Bessel}(\phi^+_1,\phi^+_2) = N^2 \int_0^1 dc \ c \ J_a(c \phi^+_1) J_a(c \phi^+_2)
\end{eqnarray}
in the limit $N \rightarrow \infty$, where
\begin{equation}
\label{phiplus}
x_1 = 1 - \frac{(\phi^+_1)^2}{2 N^2}, \ \ \ 
x_2 = 1 - \frac{(\phi^+_2)^2}{2 N^2}
\end{equation}
with $\phi^+_1, \phi^+_2 > 0$ fixed. This asymptotic form $K_a^{Bessel}(\phi^+_1,\phi^+_2)$ is 
called the Bessel kernel, as $J_a(x)$ is the Bessel function. The asymptotic correlation functions 
of this 1D model have determinant forms in terms of the Bessel kernel.
\par
The Chebyshev polynomials are special cases of the Jacobi polynomials. 
In particular the cases $(a,b) = (-1/2,-1/2)$ and $(1/2,1/2)$ are the Chebyshev 
polynomials of the first and second kind, respectively. Now we put $a = -1/2$ in 
the Bessel kernel, we obtain
\begin{equation}
\label{bessel1}
K_{-1/2}^{Bessel}(\phi^+_1,\phi^+_2) = \frac{N^2}{\pi} \frac{1}{\sqrt{\phi^+_1 \phi^+_2}} 
\left( \frac{\sin(\phi^+_1 - \phi^+_2)}{\phi^+_1 -  \phi^+_2} + 
         \frac{\sin(\phi^+_1 + \phi^+_2)}{\phi^+_1 +  \phi^+_2} \right). \nonumber \\
\end{equation}
This special case of Bessel kernel can be derived from a similar 
asymptotic form (\ref{k1limit-sine}) for $K_N^{(1)}(z_1,z_2)$ of Model I, when it is integrated 
over the transversal coordinates. 
\par
Let us explain the transversal integration in the case $\psi = 0$. Using (\ref{jacobian}) 
and the notations for (\ref{k1limit-psi=0}), 
we find
\begin{equation}
\frac{\partial(x_j,y_j)}{\partial(r_j,\theta_j)} = \frac{|1 - z_j|^2}{r_j} \sim 
\frac{| s_j | ^2}{ N^2 r_j}, \ \ N \rightarrow \infty
\end{equation}
for $j =1,2$. Therefore a factor $K_N^{(1)}(z_1,z_2) K_N^{(1)}(z_2,z_1)$ in the determinant 
forms of the correlation functions is integrated over the transversal 
coordinates $t_1,t_2$ as
\begin{eqnarray}
\label{trans}
& & \int K_N^{(1)}(z_1,z_2) K_N^{(1)}(z_2,z_1) dx_1 dy_1 dx_2 dy_2 
\nonumber \\
& = &  \int K_N^{(1)}(z_1,z_2) K_N^{(1)}(z_2,z_1)  \frac{|1 - z_1^2 |^2}{r_1}  
\frac{| 1 - z_2^2  |^2}{r_2}dr_1 d\theta_1 dr_2 d\theta_2 
\nonumber \\ 
& \sim & \frac{1}{N^4}  \sum_{\pm} \int_T^u dt_1 \int _T^u dt_2 K_N^{(1)}(z_1,z_2) K_N^{(1)}(z_2,z_1)
 \frac{| s_1 s_2 |^2}{|\phi_1 \phi_2|} dx_1 dx_2, \ \ N \rightarrow \infty \nonumber \\
\end{eqnarray}
with $\phi_1,\phi_2 \neq 0$.  Here a sum
\begin{equation}
\sum_{\pm} = \sum_{\epsilon_1=\pm 1} \sum_{\epsilon_2=\pm 1}
\end{equation}
with $\phi_j = \epsilon_j |\phi_j|$ should be included in the transversal integration, because both of 
$\epsilon_j = \pm 1$ with the same $|\phi_j|$ yield the same $x_j$ ($j = 1,2$). In the derivation of the last line of 
(\ref{trans}), we use the relation 
\begin{equation}
\label{xphi}
x_j = \frac{1}{2}  \left( r_j + \frac{1}{r_j} \right) \cos \frac{\phi_j}{N}
\sim 1 +  \frac{t_j^2 - \phi_j^2}{2 N^2}, \ \ \ N \rightarrow \infty
\end{equation}
($j = 1,2$). Now we put $T = 0$,  insert the corresponding asymptotic formula (\ref{k1limit-sine}) and take the limit $u \rightarrow 0$. Then the sum $\displaystyle \sum_{\pm}$ gives a factor $4$ and we obtain
\begin{eqnarray}
& & \lim_{u \rightarrow 0} \frac{N^4}{(\pi u)^2} \int _0^u dt_1 \int_0^u dt_2 
\frac{| s_1 s_2 |^2}{|\phi_1 \phi_2|^3} \left( 
\frac{\sin(\phi_1 - \phi_2)}{\phi_1 - \phi_2} + \frac{\sin(\phi_1 + \phi_2)}{\phi_1 + \phi_2} 
\right)^2 dx_1 dx_2  \nonumber \\ 
& & = K^{(1)}_r(\phi_1,\phi_2) K^{(1)}_r(\phi_2,\phi_1) dx_1 dx_2,
\end{eqnarray}
where
\begin{equation}
\label{onekernel1}
K^{(1)}_r(\phi_1,\phi_2) = 
\frac{N^2}{\pi} \frac{1}{\sqrt{|\phi_1 \phi_2|}}  \left( 
\frac{\sin(\phi_1 - \phi_2)}{\phi_1 - \phi_2} + \frac{\sin(\phi_1 + \phi_2)}{\phi_1 + \phi_2} 
\right).
\end{equation}
In general we can similarly show
\begin{eqnarray}
& & \int K_N^{(1)}(z_1,z_2) K_N^{(1)}(z_2,z_3) \cdots K_N^{(1)}(z_{j-1}, z_j) K_N^{(1)}(z_j, z_1) 
\nonumber \\ & & \times dx_1 dy_1 dx_2 dy_2 \cdots dx_j dy_j\nonumber \\ 
& \sim &  K^{(1)}_r(\phi_1,\phi_2) K^{(1)}_r(\phi_2,\phi_3) \cdots K^{(1)}_r(\phi_{j-1}, \phi_j) 
K^{(1)}_r(\phi_j, \phi_1) 
dx_1 dx_2 \cdots dx_j, \nonumber \\ & & j = 2,3,\cdots,\kappa
\end{eqnarray}
and
\begin{equation}
\int K_N^{(1)}(z_1,z_1) dx_1 dy_1 \sim K^{(1)}_r(\phi_1,\phi_1) dx_1
\end{equation} 
for integrals over transversal coordinates with an integer $\kappa \geq 2$ .  As a determinant can be defined as a sum over permutations and a permutation has a cycle decomposition, it follows that
\begin{eqnarray}
& & \int  \det[ K_N^{(1)}(z_j,z_\ell)]_{j,\ell = 1,2,\cdots,k} 
dx_1 dy_1 dx_2 dy_2 \cdots dx_k dy_k \nonumber \\ 
& \sim &  \det[ K^{(1)}_r(\phi_j,\phi_\ell) ]_{j,\ell = 1,2, \cdots,k} dx_1 dx_2 \cdots dx_k
\end{eqnarray}
for the transversal integration of the determinant formula (\ref{rhodet}). Here  $z_j = x_j + i y_j$, 
\begin{equation}
x_j = \frac{1}{2}  \left( r_j + \frac{1}{r_j} \right) \cos \theta_j, \  \ \ 
y_j = \frac{1}{2}  \left( r_j - \frac{1}{r_j} \right) \sin \theta_j
\end{equation}
with $\theta_j = \phi_j/N$ ($r_j > 1$, $\phi_j$ real) for $j = 1,2,\cdots,k$. Therefore $K^{(1)}_r(\phi_j,\phi_\ell)$ 
with $j,\ell=1,2,\cdots,k$ gives the 1D kernel function after the transversal integration. Since
\begin{equation}
K^{(1)}_r(\phi_j,\phi_\ell) = K^{(1)}_r(-\phi_j,\phi_\ell) = K^{(1)}_r(\phi_j,-\phi_\ell) = K^{(1)}_r(-\phi_j,-\phi_\ell),
\end{equation}
we find $K^{(1)}_r(\phi_j,\phi_\ell) = K^{(1)}_r(|\phi_j|,|\phi_\ell|)$. Because of (\ref{phiplus}) and (\ref{xphi}), 
$\phi^+_j$ can be equated with $|\phi_j|$ in the limit $u \rightarrow 0$,  so that the 1D kernel functions 
(\ref{bessel1}) and (\ref{onekernel1}) satisfy
\begin{equation}
K_{-1/2}^{Bessel}(\phi^+_j,\phi^+_\ell) \sim K^{(1)}_r(|\phi_j|,|\phi_\ell|) = K^{(1)}_r(\phi_j,\phi_\ell),
\end{equation}
as expected.
\par
Let us next put $a = 1/2$ in the Bessel kernel (\ref{bessel-kernel}). Then we find
\begin{equation}
\label{bessel2}
K_{1/2}^{Bessel}(\phi^+_1,\phi^+_2) = \frac{N^2}{\pi} \frac{1}{\sqrt{\phi^+_1 \phi^+_2}} 
\left( \frac{\sin(\phi^+_1 - \phi^+_2)}{\phi^+_1 -  \phi^+_2} -
         \frac{\sin(\phi^+_1 + \phi^+_2)}{\phi^+_1 +  \phi^+_2} \right). \nonumber \\ 
\end{equation}
As before, we can derive this  special case of Bessel kernel from a similar 
asymptotic form (\ref{k2limit-sine}) for Model II. That is, a transversal integration of the determinant 
formula (\ref{rhodet}) leads to
\begin{eqnarray}
\label{det}
& & \int  \det[ K_N^{(2)}(z_j,z_\ell)]_{j,\ell = 1,2,\cdots,k} 
dx_1 dy_1 dx_2 dy_2 \cdots dx_k dy_k \nonumber \\ 
& \sim &  \det[ K^{(2)}_r(\phi_j,\phi_\ell) ]_{j,\ell = 1,2, \cdots,k} dx_1 dx_2 \cdots dx_k, 
\end{eqnarray}
where
\begin{equation}
\label{onekernel2}
K^{(2)}_r(\phi_j,\phi_\ell) = 
\frac{N^2}{\pi} \frac{1}{\sqrt{|\phi_j \phi_\ell|}}  \left( 
\frac{\sin(\phi_j - \phi_\ell)}{\phi_j - \phi_\ell} - \frac{\sin(\phi_j + \phi_\ell)}{\phi_j + \phi_\ell}
\right)
\end{equation}
with $j,\ell = 1,2,\cdots,k$. Owing to a relation
\begin{equation}
K^{(2)}_r(\phi_j,\phi_\ell) = - K^{(2)}_r(-\phi_j,\phi_\ell) = - K^{(2)}_r(\phi_j,-\phi_\ell) = K^{(2)}_r(-\phi_j,-\phi_\ell),
\end{equation}
$K^{(2)}_r(\phi_j,\phi_\ell)$ can be replaced with $K^{(2)}_r(|\phi_j|,|\phi_\ell|)$ in (\ref{det}) without 
changing the determinant. As before the 1D kernel functions 
(\ref{bessel2}) and (\ref{onekernel2}) satisfy
\begin{equation}
K_{1/2}^{Bessel}(\phi^+_j,\phi^+_\ell) \sim K^{(2)}_r(|\phi_j|, |\phi_\ell |)
\end{equation}
by equating $\phi_j^+$ with $|\phi_j|$.
\par
Note that a similar argument can be applied to the sine kernel formula (\ref{sine-kernel}). 
Because both of 
\begin{equation}
\theta_j = \pm \left( \frac{\pi}{2} + \frac{\phi_j}{N} \right)
\end{equation}
give the same
\begin{equation}
x_j = \frac{1}{2}  \left( r_j + \frac{1}{r_j} \right) \cos\theta_j = \frac{1}{2}  \left( r_j + \frac{1}{r_j} \right) \cos\left( \frac{\pi}{2} + \frac{\phi_j}{N} \right)
\end{equation} 
and an asymptotic relation
\begin{equation}
x_j \sim - \frac{\phi_j}{N}, \ \ \ N \rightarrow \infty
\end{equation}
holds ($j = 1,2,\cdots,k$), we can derive a transversal integration formula
\begin{eqnarray}
& & \int  \det[ K_N^{(1)}(z_j,z_\ell)]_{j,\ell = 1,2,\cdots,k} 
dx_1 dy_1 dx_2 dy_2 \cdots dx_k dy_k \nonumber \\ 
& \sim &  \det[ K_c(\phi_j,\phi_\ell) ]_{j,\ell = 1,2, \cdots,k} dx_1 dx_2 \cdots dx_k
\end{eqnarray}
for the determinant (\ref{rhodet}). Here $K_c(\phi_j,\phi_\ell)$ is a 1D kernel function
\begin{equation}
K_c(\phi_j,\phi_\ell) =  \frac{N}{\pi} \frac{\sin(\phi_j - \phi_\ell)}{\phi_j - \phi_\ell}
\end{equation}
with $j,\ell = 1,2,\cdots,k$. This is a known result \cite{FK,NW,NS} for the Jacobi unitary ensemble.

\end{document}